\pdfoutput=1
\documentclass[%
 reprint,
superscriptaddress,
 amsmath,amssymb,
 aps,
floatfix,
]{revtex4-1}

\usepackage{graphicx}
\usepackage{dcolumn}
\usepackage{bm}

\usepackage{dsfont}
\usepackage{color,psfrag}
\usepackage[cleanup={log,aux,dvi,ps,pdf}]{} 
\usepackage{epstopdf}

\usepackage{xcolor} 

\usepackage{ifpdf}
    \ifpdf
\usepackage{tikz}
\usetikzlibrary{arrows,chains,matrix,positioning,scopes, fit, calc}
\usetikzlibrary{cd}
\usepackage{chemfig}
\fi

\usepackage{multirow}

\begin{document}

\preprint{APS/123-QED}

\title{Machine Learning of Accurate Energy-Conserving Molecular Force Fields}

\author{Stefan Chmiela}
 \affiliation{%
 Machine Learning Group, Technische Universit\"at Berlin, 10587 Berlin, Germany
}%

\author{Alexandre Tkatchenko}
 \email{alexandre.tkatchenko@uni.lu}
\affiliation{%
 Physics and Materials Science Research Unit, University of Luxembourg, L-1511 Luxembourg, Luxembourg
}%
 \affiliation{%
 Fritz-Haber-Institut der Max-Planck-Gesellschaft, 14195 Berlin, Germany
}%

\author{Huziel E. Sauceda}
 \affiliation{%
 Fritz-Haber-Institut der Max-Planck-Gesellschaft, 14195 Berlin, Germany
}%

\author{Igor Poltavsky}
 \affiliation{%
 Physics and Materials Science Research Unit, University of Luxembourg, L-1511 Luxembourg, Luxembourg
}%

\author{Kristof T. Sch\"utt}
 \affiliation{%
 Machine Learning Group, Technische Universit\"at Berlin, 10587 Berlin, Germany
}%

\author{Klaus-Robert M\"uller}%
 \email{klaus-robert.mueller@tu-berlin.de}
\affiliation{%
 Machine Learning Group, Technische Universit\"at Berlin, 10587 Berlin, Germany
}
\affiliation{%
Department of Brain and Cognitive Engineering, Korea University, Anam-dong, Seongbuk-gu, Seoul 136-713, Korea
}%
\affiliation{%
Max Planck Institute for Informatics, Stuhlsatzenhausweg, 66123 Saarbr\"ucken, Germany
}%

\date{\today}
\begin{abstract}

Using conservation of energy -- a fundamental property of closed classical and quantum mechanical systems -- we develop an efficient gradient-domain machine learning (GDML) approach to construct accurate molecular force fields using a restricted number of samples from ab initio molecular dynamics (AIMD) trajectories. The GDML implementation is able to reproduce global potential energy surfaces of intermediate-sized molecules with an accuracy of 0.3 kcal $\text{mol}^{-1}$ for energies and 1 kcal $\text{mol}^{-1}$ $\text{\AA}^{-1}$ for atomic forces using only 1000 conformational geometries for training. We demonstrate this accuracy for AIMD trajectories of molecules, including benzene, toluene, naphthalene, ethanol, uracil, and aspirin. The challenge of constructing conservative force fields is accomplished in our work by learning in a Hilbert space of vector-valued functions that obey the law of energy conservation. The GDML approach enables quantitative molecular dynamics simulations for molecules at a fraction of cost of explicit AIMD calculations, thereby allowing the construction of efficient force fields with the accuracy and transferability of high-level ab initio methods.
\end{abstract}

\pacs{Valid PACS appear here}
\maketitle


\section{Introduction}

\begin{figure*}[ht]
\centering
\includegraphics[width=\textwidth]{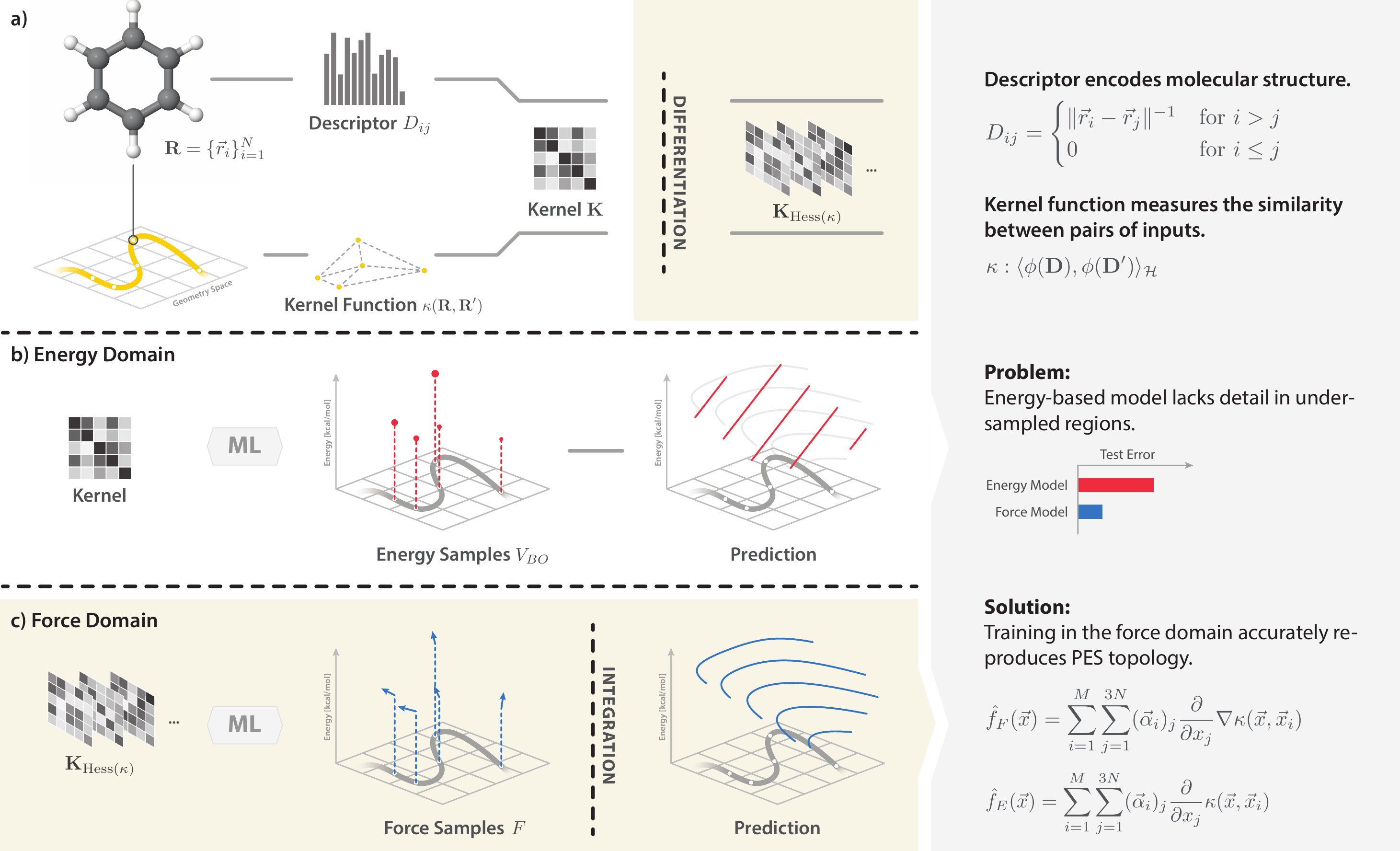}
\caption{The construction of ML models: First, reference data from an MD trajectory are sampled. (a) The geometry of each molecule is encoded in a descriptor. This representation introduces elementary transformational invariances of energy and constitutes the first part of the prior. A kernel function then relates all descriptors to form the kernel matrix -- the second part of the prior. The kernel function encodes similarity between data points. Our particular choice makes only weak assumptions: It limits the frequency spectrum of the resulting model and adds the energy conservation constraint. Hess, Hessian. (c) These general priors are sufficient to reproduce good estimates from a restricted number of force samples. (b) A comparable energy model is not able to reproduce the PES to the same level of detail.}
\label{fig:main_figure}
\end{figure*}

Within the Born-Oppenheimer (BO) approximation, predictive simulations of properties and functions of molecular systems require an accurate description of the global potential energy hypersurface $V_{\rm{BO}}(\vec{r}_1,\vec{r}_2,\dotsc,\vec{r}_N)$, where $\vec{r}_i$ indicates the nuclear Cartesian coordinates. Although $V_{\rm{BO}}$ could, in principle, be obtained on the fly using explicit ab initio calculations, more efficient approaches that can access the long time scales are required to understand relevant phenomena in large molecular systems. A plethora of classical mechanistic approximations to $V_{\rm{BO}}$ have been constructed, in which the parameters are typically fitted to a small set of ab initio calculations or experimental data. Unfortunately, these classical approximations may suffer from the lack of transferability and can yield accurate results only close to the conditions (geometries) they have been fitted to. Alternatively, sophisticated machine learning (ML) approaches that can accurately reproduce the global potential energy surface (PES) for elemental materials~\cite{Behler2007, doi:10.1063/1.2746232, Bartok2010, Behler2011, Behler2011a, doi:10.1063/1.4712397, Bartok2013, Bartok2015, C6CP00415F} and small molecules ~\cite{Rupp2012, Montavon2013a, Hansen2013, Hansen2015, Rupp2015, Ramprasad2015, DBLP:journals/corr/HirnPM15} have been recently developed (see Fig.~\ref{fig:main_figure}, a and b)~\cite{behler2016}. Although potentially very promising, one particular challenge for direct ML fitting of molecular PES is the large amount of data necessary to obtain an accurate model. Often, many thousands or even millions of atomic configurations are used as training data for ML models. This results in nontransparent models, which are difficult to analyze and may break consistency~\cite{DeVita2015} between energies and forces.

A fundamental property that any force field ${\mathbf F}_i(\vec{r}_1,\vec{r}_2,\dotsc,\vec{r}_N)$ must satisfy is the conservation of total energy, which implies that ${\mathbf F}_i(\vec{r}_1,\vec{r}_2,\dotsc,\vec{r}_N) = -\nabla_{\vec{r}_i} V(\vec{r}_1,\vec{r}_2,\dotsc,\vec{r}_N)$. Any classical mechanistic expressions for the potential energy (also denoted as classical force field) or analytically derivable ML approaches trained on energies satisfy energy conservation by construction. However, even if conservation of energy is satisfied implicitly within an approximation, this does not imply that the model will be able to accurately follow the trajectory of the true ab initio potential, which was used to fit the force field. In particular, small energy/force inconsistencies between the force field model and ab initio calculations can lead to unforeseen artifacts in the PES topology, such as spurious critical points that can give rise to incorrect molecular dynamics (MD) trajectories. Another fundamental problem is that classical and ML force fields focusing on energy as the main observable have to assume atomic energy additivity -- an approximation that is hard to justify from quantum mechanics.

Here, we present a robust solution to these challenges by constructing an explicitly conservative ML force field, which uses exclusively atomic gradient information in lieu of atomic (or total) energies. In this manner, with any number of data samples, the proposed model fulfills energy conservation by construction. Obviously, the developed ML force field can be coupled to a heat bath, making the full system (molecule and bath) non-energy-conserving.

We remark that atomic forces are true quantum-mechanical observables within the BO approximation by virtue of the Hellmann-Feynman theorem. The energy of a molecular system is recovered by analytic integration of the force-field kernel (see Fig.~\ref{fig:main_figure}c). We demonstrate that our gradient-domain machine learning (GDML) approach is able to accurately reproduce global PESs of intermediate-sized molecules within 0.3 kcal $\text{mol}^{-1}$ for energies and 1 kcal $\text{mol}^{-1}$ $\text{\AA}^{-1}$ for atomic forces relative to the reference data. This accuracy is achieved when using less than 1000 training geometries to construct the GDML model and using energy conservation to avoid overfitting and artifacts. Hence, the GDML approach paves the way for efficient and precise MD simulations with PESs that are obtained with arbitrary high-level quantum- chemical approaches. We demonstrate the accuracy of GDML by computing AIMD-quality thermodynamic observables using path-integral MD (PIMD) for eight organic molecules with up to 21 atoms and four chemical elements. Although we use density functional theory (DFT) calculations as reference in this development work, it is possible to use any higher-level quantum-chemical reference data. With state-of-the-art quantum chemistry codes running on current high-performance computers, it is possible to generate accurate reference data for molecules with a few dozen atoms. Here, we focus on intramolecular forces in small- and medium-sized molecules. However, in the future, the GDML model should be combined with an accurate model for intermolecular forces to enable predictive simulations of condensed molecular systems. Widely used classical mechanistic force fields are based on simple harmonic terms for intramolecular degrees of freedom. Our GDML model correctly treats anharmonicities by using no assumptions whatsoever on the analytic form on the interatomic potential energy functions within molecules.

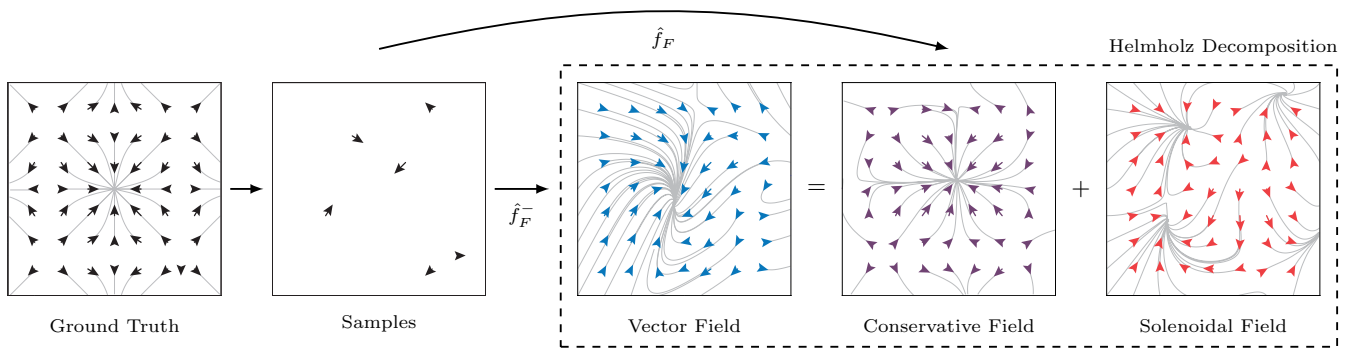
\begin{figure*}[!ht]
\hspace*{-0.31cm}
\begin{tikzpicture}
\matrix (m) [matrix of math nodes, row sep=0, column sep=1.4em, font=\scriptsize]
{
\node[label={[xshift=0.0cm, yshift=-3.55cm]:\text{Ground Truth}}](truth){\includegraphics[scale=0.235]{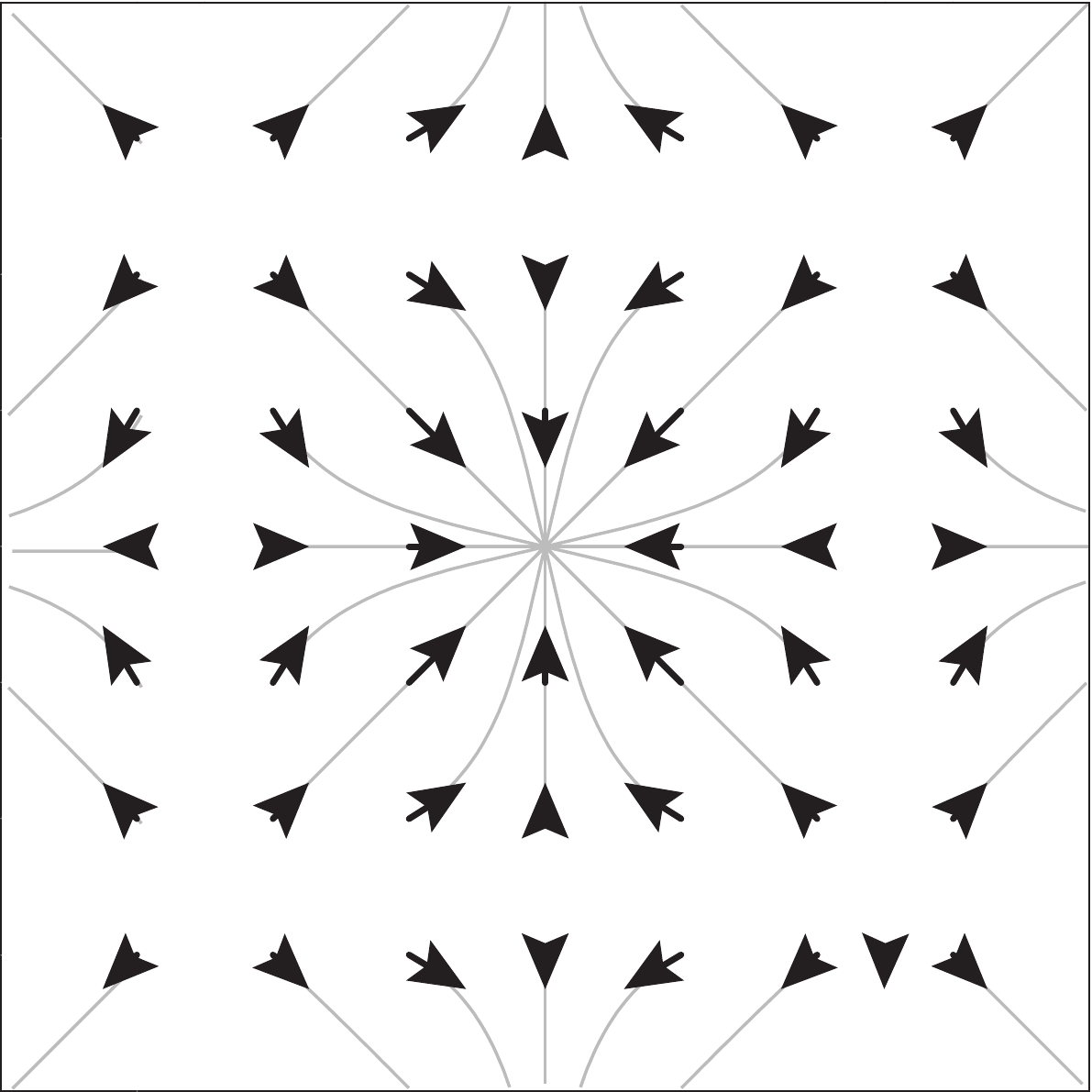}};
& \node[label={[xshift=0.0cm, yshift=-3.55cm]:\text{Samples}}](samples){\includegraphics[scale=0.235]{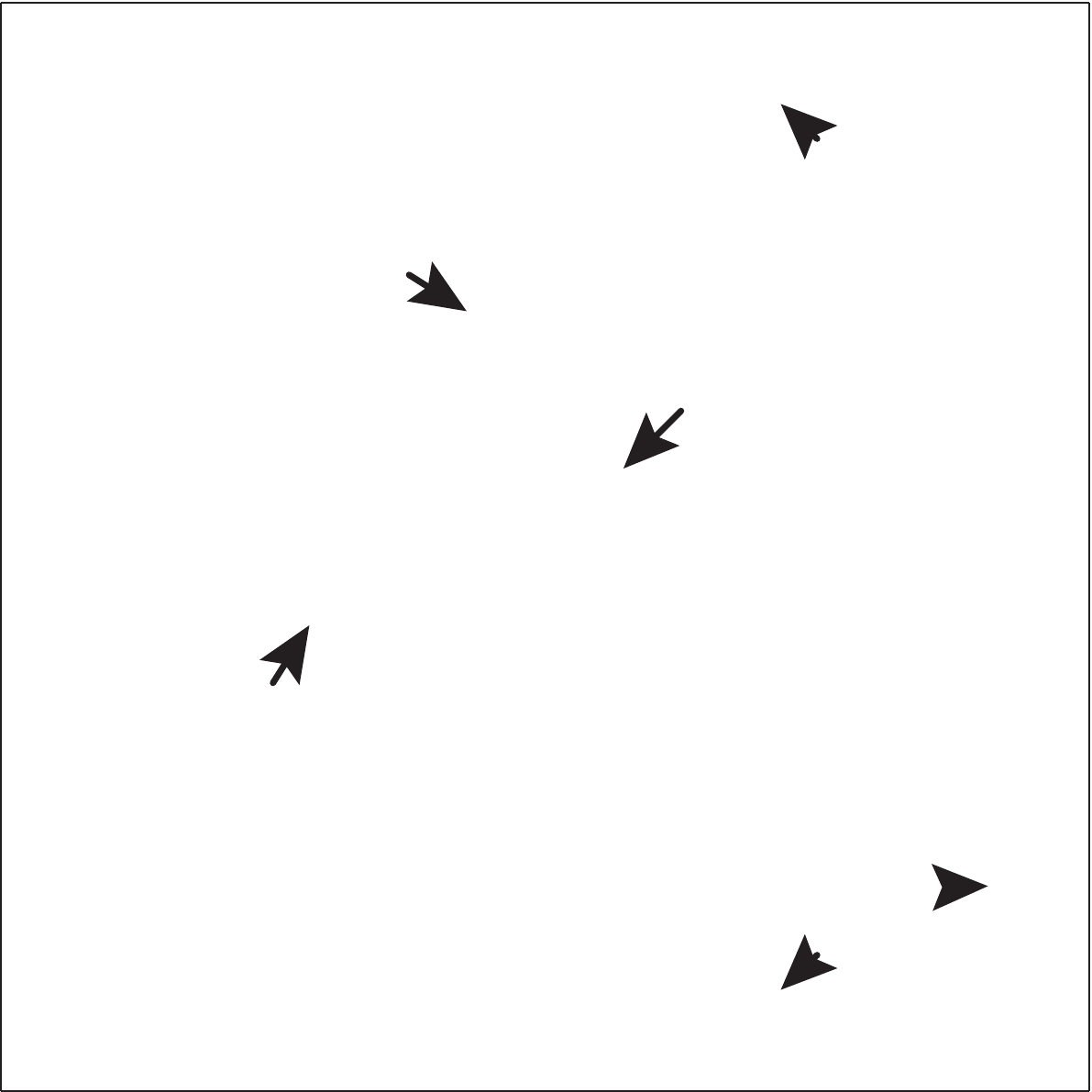}};
&[4ex] \node[label={[xshift=0.0cm, yshift=-3.55cm]:\text{Vector Field}}](naive){\includegraphics[scale=0.235]{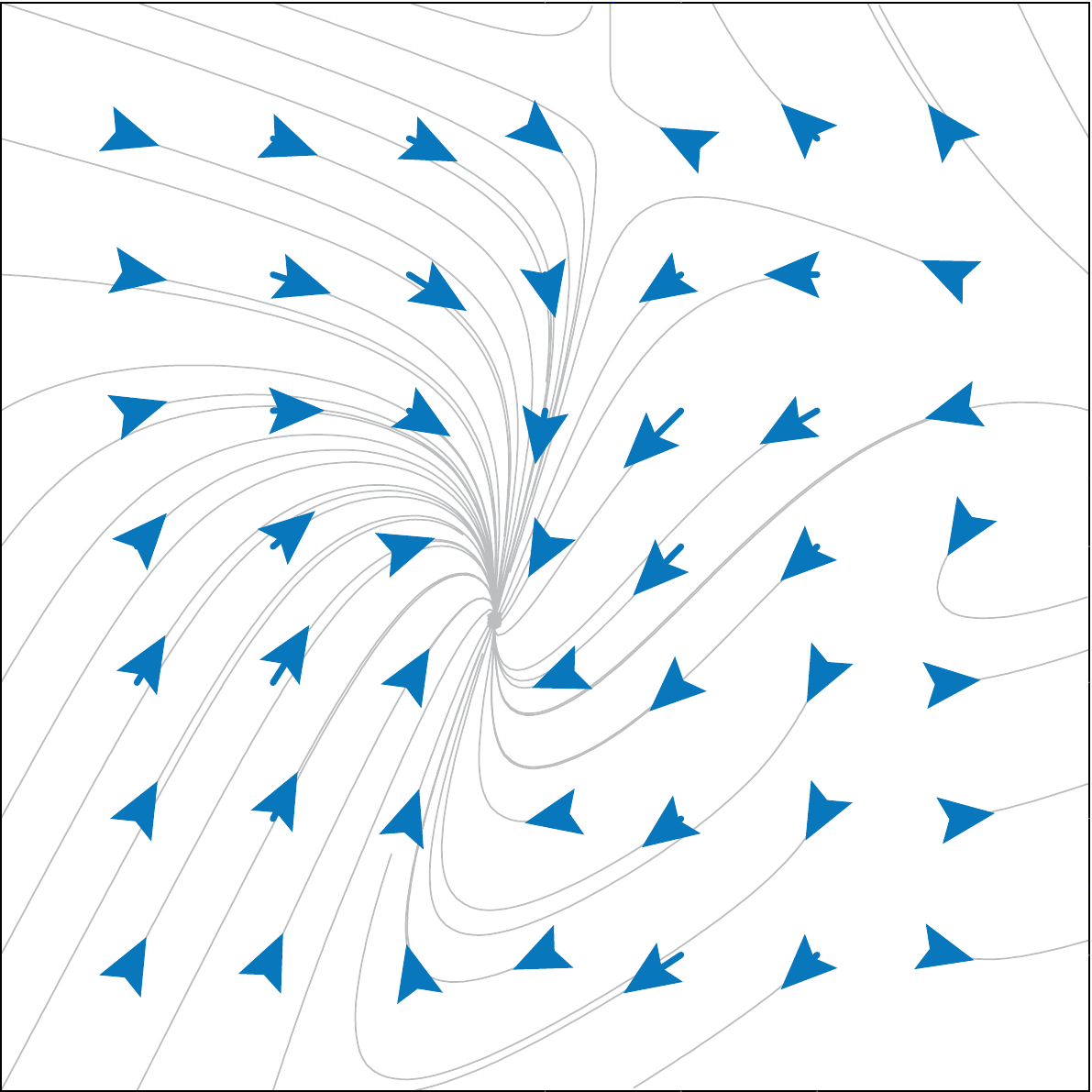}};  \coordinate (l1) at (0,-0.57);
& \node[label={[xshift=0.0cm, yshift=-3.55cm]:\text{Conservative Field}}](curl_free){\includegraphics[scale=0.235]{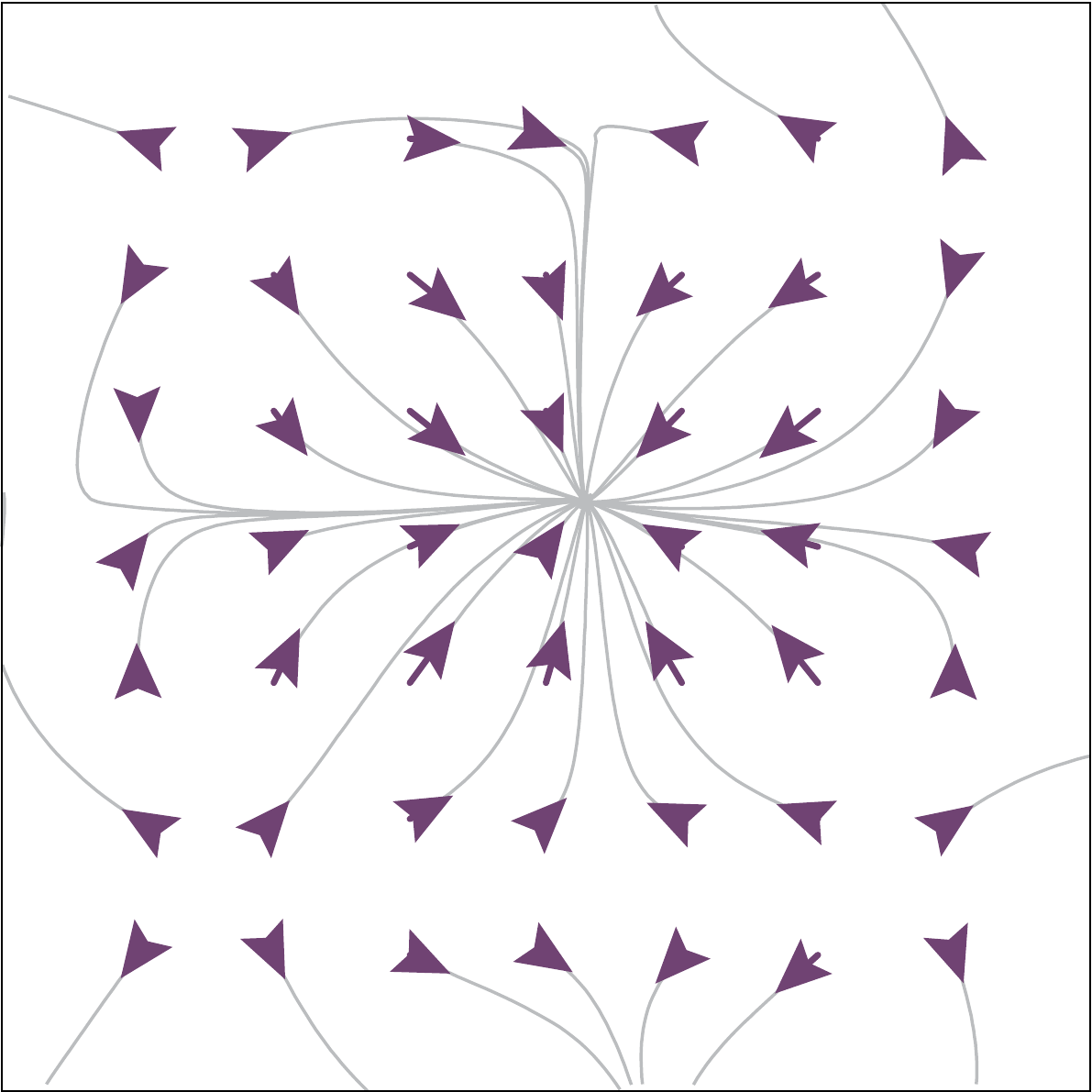}};
& \node[label={[xshift=0.0cm, yshift=-3.55cm]:\text{Solenoidal Field}}](div_free){\includegraphics[scale=0.235]{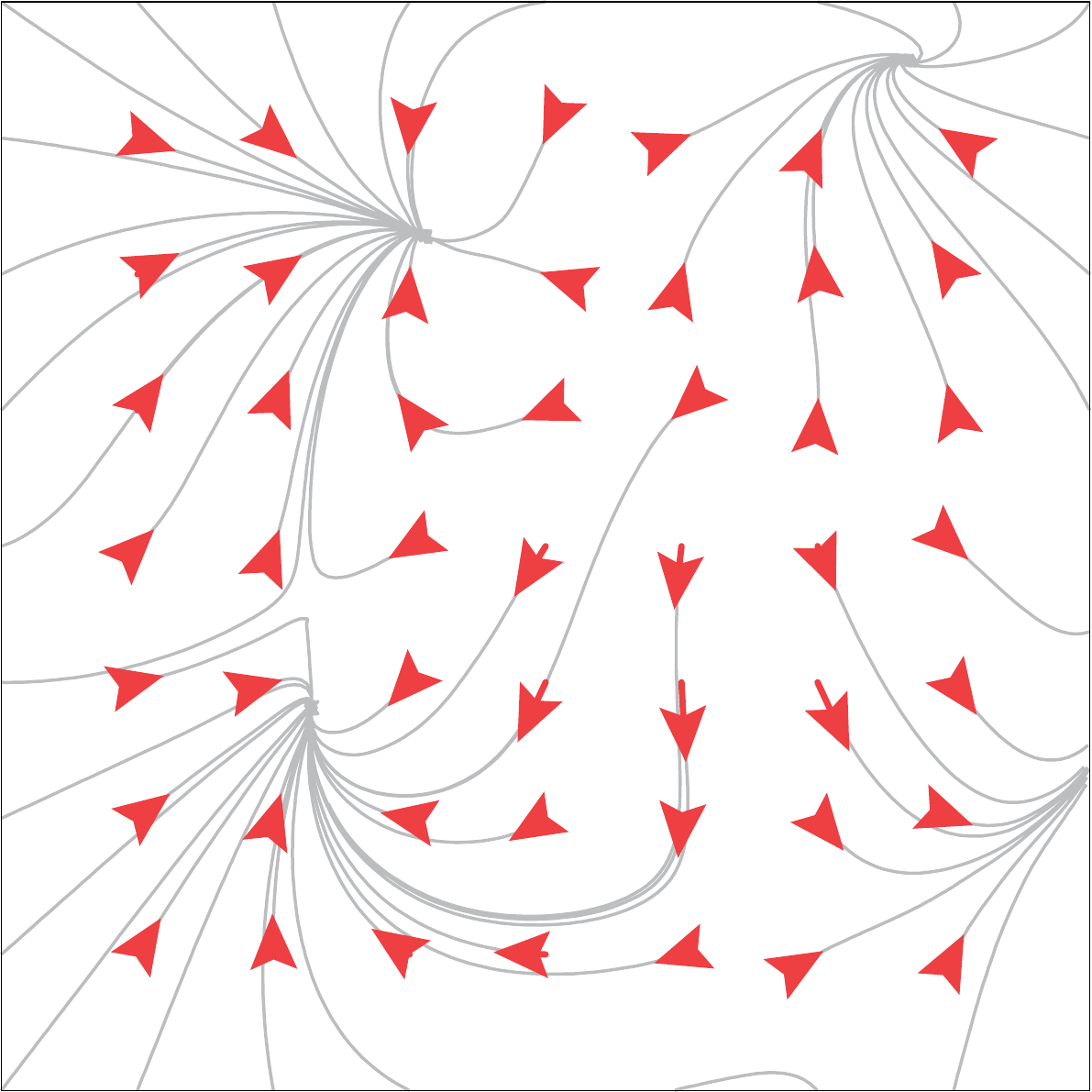}}; \\ 
};
\path[->]
(truth) edge [-latex, line width=0.25mm] (samples)
(samples) edge [-latex, line width=0.25mm]  node[yshift=-0.35cm, font=\scriptsize]{$\hat{f}^{-}_F$} ($(naive.west)+(-0.8em,0)$)
(naive)edge [-{Stealth}, draw=none, line width=0.25mm] node{$ = $} (curl_free)
([yshift=1em]samples.north)edge [-latex, bend right=-13, line width=0.25mm]  node[yshift=-0.35cm, font=\scriptsize]{$\hat{f}_F$}  ([yshift=1em]curl_free.north)
(curl_free)edge[-{Stealth}, draw=none] node{$ + $}  (div_free);
\node[label={[xshift=3.65cm, yshift=0.0cm, font=\scriptsize]:\text{Helmholz Decomposition}}][draw, line width=0.25mm, dashed,fit=(naive) (curl_free) (div_free) (l1)] {};
\end{tikzpicture}
\caption{Modeling the true vector field (leftmost subfigure) based on a small number of vector samples. With GDML, a conservative vector field estimate $\hat{f}_F$ is obtained directly. A na\"{\i}ve estimator $\hat{f}^-_F$ with independent predictions for each element of the output vector is not capable of imposing energy conservation constraints. We perform a Helmholtz decomposition of this nonconservative vector field to show the error component that violates the law of energy conservation. This is the portion of the overall prediction error that was avoided with GDML because of the addition of the energy conservation constraint.}
\label{fig:ex2_schema}
\end{figure*}
%


\section{Method}

The GDML approach explicitly constructs an energy-conserving force field, avoiding the application of the noise-amplifying derivative operator to a parameterized potential energy model (see the Supplementary Materials for details). This can be achieved by directly learning the functional relationship

\begin{equation}
\hat{f}_\text{F}: (\vec{r}_1,\vec{r}_2,\dotsc,\vec{r}_N)_{i} \xrightarrow{\text{ML}} {\mathbf {F}}_i
\end{equation}
between atomic coordinates and interatomic forces, instead of computing the gradient of the PES (see Fig.~\ref{fig:main_figure}, c and b). This requires constraining the solution space of all arbitrary vector fields to the subset of energy-conserving gradient fields. The PES can be obtained through direct integration of $\hat{f}_\text{F}$ up to an additive constant.

To construct $\hat{f}_\text{F}$, we used a generalization of the commonly used kernel ridge regression technique for structured vector fields (see the Supplementary Materials for details)~\cite{Micchelli2005, Caponnetto2008, Sindhwani1964}. GDML solves the normal equation of the ridge estimator in the gradient domain using the Hessian matrix of a kernel as the covariance structure. It maps to all partial forces of a molecule simultaneously (see Fig.~\ref{fig:main_figure}a)
\begin{equation}
\left(\mathbf{K}_{\text{Hess}(\kappa)} + \lambda \mathbb{I}\right) \vec{\alpha}= \nabla V_{BO} = -\mathbf{F}
\end{equation}

We resorted to the extensive body of research on suitable kernels and descriptors for the energy prediction task~\cite{Rupp2012,Hansen2015,behler2016}. For our application, we considered a subclass from the parametric Mat\'{e}rn family~\cite{matern1986spatial, Gradshteyn2007, Gneiting2010} of (isotropic) kernel functions
\begin{equation}
  \begin{aligned}
\kappa: C_{v = n + \frac{1}{2}}(d) &= \exp\left(-\frac{\sqrt{2v}d}{\sigma} \right) P_n(d) \text{,}\\
P_n(d) &= \sum^n_{k=0} \frac{(n+k)!}{(2n)!} {n \choose k} \left( \frac{2\sqrt{2v} d}{\sigma} \right)^{n-k}
  \end{aligned}
\end{equation}
where $d = \|\vec{x}- \vec{x}'\|$ is the Euclidean distance between two molecule descriptors. It can be regarded as a generalization of the universal Gaussian kernel with an additional smoothness parameter $n$. Our parameterization $n=2$ resembles the Laplacian kernel, as suggested by Hansen et al.~\cite{Hansen2015}, while being sufficiently differentiable.

To disambiguate Cartesian geometries that are physically equivalent, we use an input descriptor derived from the Coulomb matrix (see the Supplementary Materials for details)~\cite{Rupp2012}.

The trained force field estimator collects the contributions of the partial derivatives $3N$ of all training points $M$ to compile the prediction. It takes the form
\begin{equation}
\hat{f}_F(\vec{x}) = \sum^M_{i=1} \sum^{3N}_{j=1} (\vec{\alpha}_{i})_j \frac{\partial}{\partial x_{j}} \nabla \kappa(\vec{x},\vec{x}_i)
\label{eq:force_model}
\end{equation}
and a corresponding energy predictor is obtained by integrating $\hat{f}_F(\vec{x})$ with respect to the Cartesian geometry. Because the trained model is a (fixed) linear combination of kernel functions, integration only affects the kernel function itself. The expression
\begin{equation}
\hat{f}_E(\vec{x}) = \sum^M_{i=1} \sum^{3N}_{j=1} (\vec{\alpha}_{i})_j \frac{\partial}{\partial x_{j}}\kappa(\vec{x},\vec{x}_i)
\label{eq:energy_model} 
\end{equation}
for the energy predictor is therefore neither problem-specific nor does it require retraining.

We remark that our PES model is global in the sense that each molecular descriptor is considered as a whole entity, bypassing the need for arbitrary partitioning of energy into atomic contributions. This allows the GDML framework to capture chemical and long-range interactions. Obviously, long-range electrostatic and van der Waals interactions that fall within the error of the GDML model will have to be incorporated with explicit physical models. Other approaches that use ML to fit PESs such as Gaussian approximation potentials~\cite{Bartok2010, Bartok2015} have been proposed. However, these approaches consider an explicit localization of the contribution of individual atoms to the total energy. The total energy is expressed as a linear combination of local environments characterized by a descriptor that acts as a non-unique partitioning function to the total energy. Training on force samples similarly requires the evaluation of kernel derivatives, but w.r.t. those local environments. Although any partitioning of the total energy is arbitrary, our molecular total energy is physically meaningful in that it is related to the atomic force, thus being a measure for the deflection of every atom from its ground state.

We first demonstrate the impact of the energy conservation constraint on a toy model that can be easily visualized. A nonconservative force model $\hat{f}^-_F$ was trained alongside our GDML model $\hat{f}_F$ on a synthetic potential defined by a two-dimensional harmonic oscillator using the same samples, descriptor, and kernel.

We were interested in a qualitative assessment of the prediction error that is introduced as a direct result of violating the law of energy conservation.
For this, we uniquely decomposed our na\"{\i}ve estimate
\begin{equation}
\hat{f}^-_F = - \nabla E + \nabla \times A
\end{equation}
into a sum of a curl-free (conservative) and a divergence-free (solenoidal) vector field, according to the Helmholtz theorem (see Fig.~\ref{fig:ex2_schema})~\cite{Helmholz1858}. This was achieved by subsampling $\hat{f}^-_F$ on a regular grid and numerically projecting it onto the closest conservative vector field by solving Poisson's equation~\cite{Press:2007:NRE:1403886}.
\begin{equation}
-\nabla^2 E \overset{!}{=} \nabla \hat{f}^-_F
\label{eq:poission_equation}
\end{equation}
with Neumann boundary conditions. The remaining solenoidal field represents the systematic error made by the na\"{\i}ve estimator. Other than in this example, our GDML approach directly estimates the conservative vector field and does not require a costly numerical projection on a dense grid of regularly spaced samples.

\section{Results}

We now proceed to evaluate the performance of the GDML approach by learning and then predicting AIMD trajectories for molecules, including benzene, uracil, naphthalene, aspirin, salicylic acid, malonaldehyde, ethanol, and toluene (see table S1 for details of these molecular data sets). These data sets range in size from 150 k to nearly 1 M conformational geometries with a resolution of 0.5 fs, although only a drastically reduced subset is necessary to train our energy and GDML predictors. The molecules have different sizes, and the molecular PESs exhibit different levels of complexity. The energy range across all data points within a set spans from 20 to 48 kcal $\text{mol}^{-1}$. Force components range from 266 to 570 kcal $\text{mol}^{-1}$ $\text{\AA}^{-1}$. The total energy and force labels for each data set were computed using the PBE+vdW-TS electronic structure method~\cite{PBE,TS}.

\begin{figure} 
  \begin{minipage}{\columnwidth} 
     \centering 
     \includegraphics[width=\columnwidth]{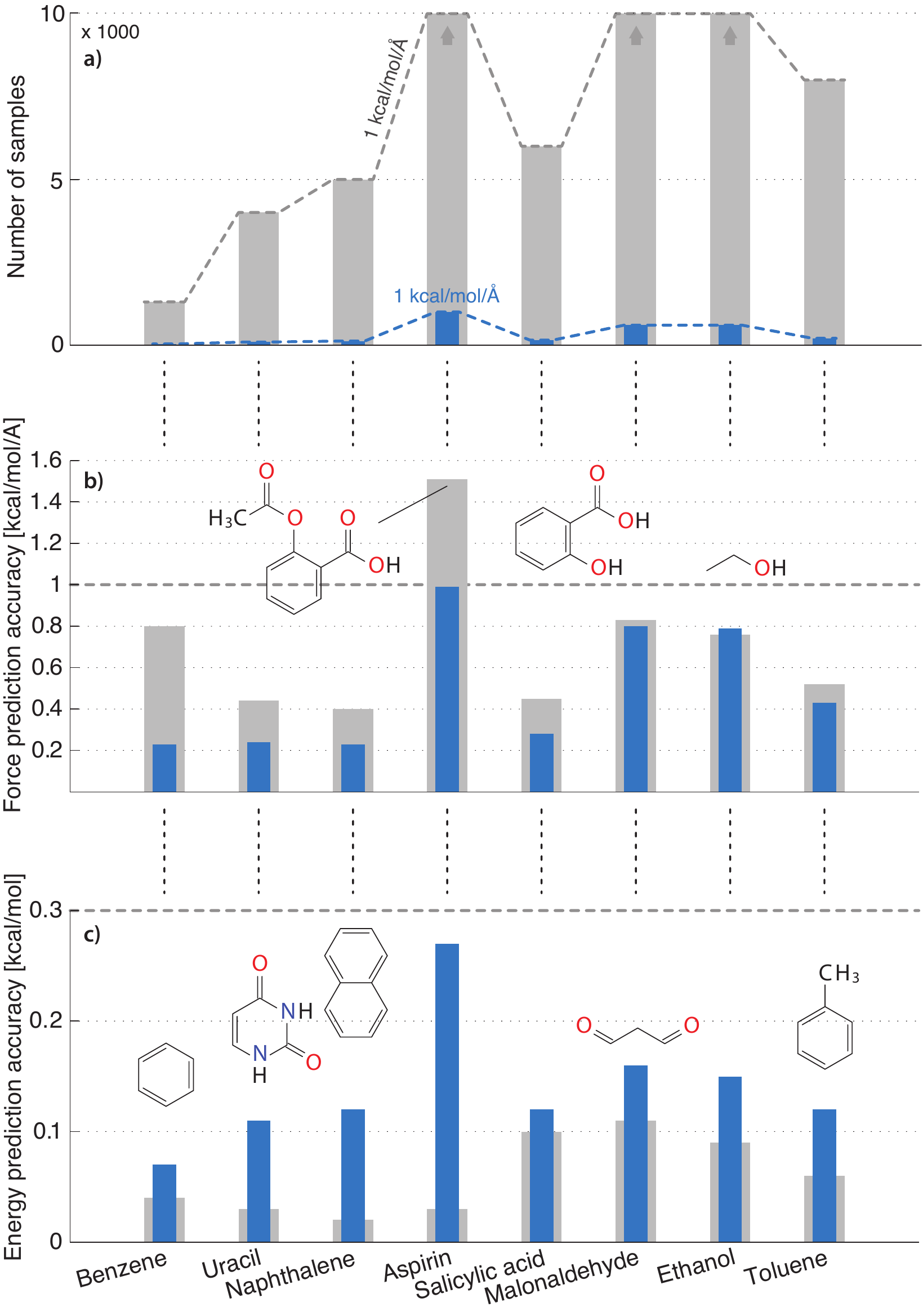} 
  \end{minipage} 
\caption{Efficiency of GDML predictor versus a model that has been trained on energies. (a) Required number of samples for a force prediction performance of MAE (1 kcal $\text{mol}^{-1}$ $\text{\AA}^{-1}$) with the energy-based model (gray) and GDML (blue). The
energy-based model was not able to achieve the targeted performance with the maximum number of 63,000 samples for aspirin. (b) Force prediction errors for the converged models (same number of partial derivative samples and energy samples). (c) Energy prediction errors for the converged models. All reported prediction errors have been estimated via cross-validation.}
  
\label{fig:bars}
\end{figure}

The GDML prediction results are contrasted with the output of a model that has been trained on energies. Both models use the same kernel and descriptor, but the hyperparameter search was performed individually to ensure optimal model selection. The GDML model for each data set was trained on $\sim$1000 geometries, sampled uniformly according to the MD@DFT trajectory energy distribution. For the energy model, we multiplied this amount by the number of atoms in one molecule times its three spatial degrees of freedom. This configuration yields equal kernel sizes for both models and therefore equal levels of complexity in terms of the optimization problem. We compare the models on the basis of the required number of samples (Fig.~\ref{fig:bars}a) to achieve a force prediction accuracy of 1 kcal $\text{mol}^{-1}$ $\text{\AA}^{-1}$. Furthermore, the prediction accuracy of the force and energy estimates for fully converged models (w.r.t. number of samples) (Fig.~\ref{fig:bars}, b and c) are judged on the basis of the mean absolute error (MAE) and root mean square error performance measures.

\begin{figure}[b]
  \begin{minipage}{\columnwidth} 
     \centering 
     \includegraphics[width=\columnwidth]{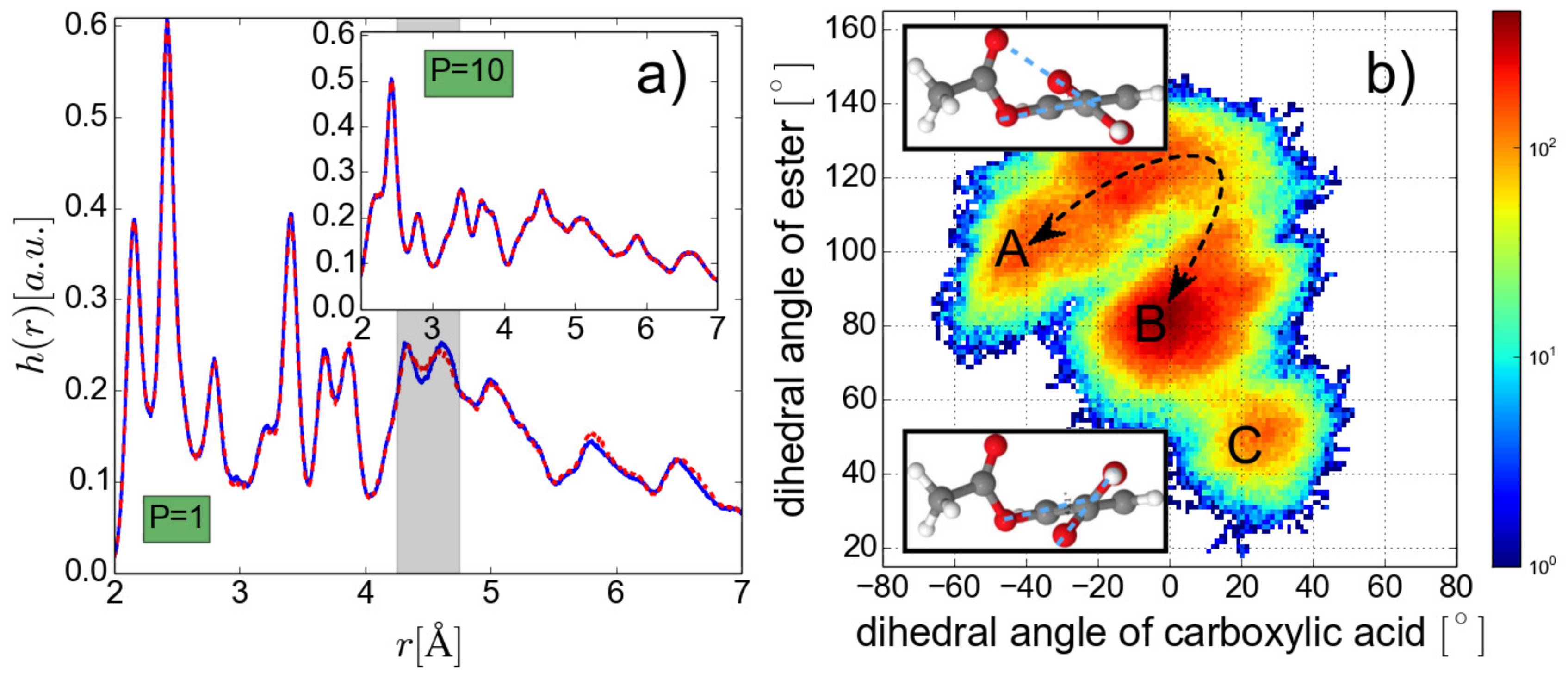} 
  \end{minipage} 
\caption{Results of classical and PIMD simulations. The recently developed estimators based on perturbation theory were used to evaluate structural and electronic observables~\cite{Poltavsky2016}. (a) Comparison of the interatomic distance distributions, $h(r) = \left\langle \frac{2}{N(N-1)} \sum^N_{i<j} \delta\left(r-\|\vec{r}_i-\vec{r}_j\|\right) \right\rangle_{P,t}$, obtained from GDML (blue line) and DFT (dashed red line) with classical MD (main frame), and PIMD (inset). a.u., arbitrary units. (b) Probability distribution of the dihedral angles (corresponding to carboxylic acid and ester functional groups) using a 20 ps time interval from a total PIMD trajectory of 200 ps.}
\label{fig:figPIMD}
\end{figure}

It can be seen in Fig.~\ref{fig:bars}a that the GDML model achieves a force accuracy of 1 kcal $\text{mol}^{-1}$ $\text{\AA}^{-1}$ using only $\sim$1000 samples from different data sets. Conversely, a pure energy-based model would require up to two orders of magnitude more samples to achieve a similar accuracy. The superior performance of the GDML model cannot be simply attributed to the greater information content of force samples. We compare our results to those of a na\"{\i}ve force model along the lines of the toy example shown in Fig.~\ref{fig:ex2_schema} (see tables S1 and S3 for details on the prediction accuracy of both models). The na\"{\i}ve force model is nonconservative but identical to the GDML model in all other aspects. Note that its performance deteriorates significantly on all data sets compared to the full GDML model (see the Supplementary Materials for details). We note here that we used DFT calculations, but any other high-level quantum chemistry approach could have been used to calculate forces for 1000 conformational geometries. This allows AIMD simulations to be carried out at the speed of ML models with the accuracy of correlated quantum chemistry calculations.

It is noticeable that the GDML model at convergence (w.r.t. number of samples) yields higher accuracy for forces than an equivalent energy-based model (see Fig.~\ref{fig:bars}b). Here, we should remark that the energy- based model trained on a very large data set can reduce the energy error to below 0.1 kcal $\text{mol}^{-1}$, whereas the GDML energy error remains at 0.2 kcal $\text{mol}^{-1}$ for $\sim$1000 training samples (see Fig.~\ref{fig:bars}c). However, these errors are already significantly below thermal fluctuations ($k_B T$) at room temperature ($\sim$0.6 kcal $\text{mol}^{-1}$), indicating that the GDML model provides an excellent description of both energies and forces, fully preserves their consistency, and reduces the complexity of the ML model. These are all desirable features of models that combine rigorous physical laws with the power of data-driven machines.

The ultimate test of any force field model is to establish its aptitude to predict statistical averages and fluctuations using MD simulations. The quantitative performance of the GDML model is demonstrated in Fig.~\ref{fig:figPIMD} for classical and quantum MD simulations of aspirin at T = 300 K. Figure~\ref{fig:figPIMD} a shows a comparison of interatomic distance distributions, $h(r)$, from MD@DFT and MD@GDML. Overall, we observe a quantitative agreement in $h(r)$ between DFT and GDML simulations. The small differences in the distance range between 4.3 and 4.7 {\AA} result from slightly higher energy barriers of the GDML model in the pathway from A to B corresponding to the collective motions of the carboxylic acid and ester groups in aspirin. These differences vanish once the quantum nature of the nuclei is introduced in the PIMD simulations~\cite{Ceriotti20141019}. In addition, long-time scale simulations are required to completely understand the dynamics of molecular systems. Figure 4B shows the probability distribution of the fluctuations of dihedral angles of carboxylic acid and ester groups in aspirin. This plot shows the existence of two main metastable configurations A and B and a short-lived configuration C, illustrating the nontrivial dynamics captured by the GDML model. Finally, we remark that a similarly good performance as for aspirin is also observed for the other seven molecules shown in Fig.~\ref{fig:bars}. The efficiency of the GDML model (which is three orders of magnitude faster than DFT) should enable long-time scale PIMD simulations to obtain converged thermodynamic properties of intermediate-sized molecules with the accuracy and transferability of high-level ab initio methods.

In summary, the developed GDML model allows the construction of complex multi-dimensional PES by combining rigorous physical laws with data-driven ML techniques. In addition to the presented successful applications to model systems and intermediate-sized molecules, our work can be further developed in several directions, including scaling with system size and complexity, incorporating additional physical priors, describing reaction pathways and enabling seamless coupling between GDML and \emph{ab initio} calculations.

\section{Data Availability}

All data sets used in this work are available at \href{http://quantum-machine.org/datasets/}{http://quantum-machine.org/datasets/}

\section{Acknowledgements}

S.C., A.T., and K.-R.M. thank the Deutsche Forschungsgemeinschaft (project MU $987/ 20$-1) for funding this work. K.-R.M. gratefully acknowledges the BK21 program funded by the Korean National Research Foundation grant (no. 2012-005741). A.T. is funded by the European Research Council with ERC-CoG grant BeStMo. Additional support was provided by the Federal Ministry of Education and Research (BMBF) for the Berlin Big Data Center BBDC (01\textbar S14013A). Part of this research was performed while the authors were visiting the Institute for Pure and Applied Mathematics, which is supported by the NSF. 

\section{Author Information}

\subsection{Contributions}
S.C. conceived, constructed, and analyzed the GDML models. S.C., A.T., and K.-R.M. developed the theory and designed the analyses. H.E.S. and I.P. performed the DFT calculations and MD simulations. H.E.S. helped with the analyses. K.T.S. and A.T. helped with the figures. A.T., S.C., and K.-R.M. wrote the paper with contributions from other authors. All authors discussed the results and commented on the manuscript.

\subsection{Corresponding Authors}
Correspondence to Klaus-Robert M\"uller or Alexandre Tkatchenko.

\nocite{Scholkopfa, Snyder2012, Snyder2015, Scholkopf1998, Scholkopf1999, Micchelli2005, Caponnetto2008, Sindhwani1964, Muller2001, Hansen2013, Rupp2012, Ceriotti20141019, Poltavsky2016}
\bibliography{references/clean_ref}

\end{document}